 \documentclass{ws-ijmpa}

\usepackage[super,compress]{cite}
\usepackage{graphicx}
\def\be{\begin{equation}}
\def\ee{\end{equation}}
\def\bea{\begin{eqnarray}}
\def\eea{\end{eqnarray}}
\newcommand{\bp}{M_P}

\begin{document}
\markboth{Alberto Salvio}{Inflation and Reheating through an Independent Affine Connection}

%
\catchline{}{}{}{}{}
%


\title{Inflation and Reheating through an Independent Affine Connection
}

\author{Alberto Salvio\footnote{
Via della Ricerca Scientifica 1, 00133, Rome, Italy
}
}

\address{
Department of Physics, University of Rome and INFN Tor Vergata, \\
Via della Ricerca Scientifica 1, 00133, Rome, Italy \\
alberto.salvio@roma2.infn.it}

%

\maketitle


\begin{abstract}
This paper is based on a talk in which I discussed how a component of the dynamical affine connection, that is independent of the metric, can drive inflation in agreement with observations. This provides a geometrical origin for the inflaton. I also illustrated how the decays of this field, which has spin 0 and odd parity, into Higgs bosons can reheat the universe to a sufficiently high temperature.

\keywords{Alternative gravity theories; inflation; particle interactions.}
\end{abstract}



\section{Introduction}

Einstein's general relativity (GR) explains gravity geometrically: distances are measured with the metric, $g_{\mu\nu}$, and the gravitational force is determined by the (affine) connection ${\cal A}_{\mu~\sigma}^{~\,\rho}$. This elegant framework explains all gravitational observations performed so far, including the nearly-exponential accelerated expansion of the universe today, in the presence of a cosmological constant. 

It is generically accepted that another, albeit much more violent, nearly-exponential expansion occurred in the early universe (inflation). This can be due to a spin-0 field, the inflaton, with an appropriate potential, which ensures such an expansion and its end:  reheating must occur after inflation to generate all the matter we observe.
	
From a purely geometrical standpoint,  $g_{\mu\nu}$ and ${\cal A}_{\mu~\sigma}^{~\,\rho}$, unlike in GR, can be entirely  independent objects and can contain extra degrees of freedom beyond the spin-2 graviton. This generalized gravity is called ``metric-affine" (see Ref.~\citen{Baldazzi:2021kaf} for a recent discussion and references).

In this talk I  explored the question ``Can the inflaton be identified with an extra dynamical component of the connection?". The main motivation for pursuing an answer is to ascertain whether the inflaton can have a geometrical origin  too. To provide an affirmative response, one must determine if and how a spin-0 field with an appropriate potential can be extracted from ${\cal A}_{\mu~\sigma}^{~\,\rho}$, such that all latest observational bounds~\cite{Ade:2015lrj,Akrami:2018,BICEP:2021xfz} are satisfied and a viable reheating takes place after inflation.  A positive answer was found in Ref.~\citen{Salvio:2022suk} and\footnote{See also Ref.~\citen{DiMarco:2023ncs} for a more recent discussion inspired by the more general framework of Ref.~\citen{Pradisi:2022nmh}.}  in the rest of the talk, which I will report in the following sections, I described how this can be explicitly implemented.

\section{The Key Idea}

When ${\cal A}_{\mu~\sigma}^{~\,\rho}$ and $g_{\mu\nu}$ are independent there are two invariants that are linear in the curvature ({\it i.e.} the  field strength ${\cal A}_{\mu~\sigma}^{~\,\rho}$)
$$ {\cal R}_{\mu\nu~~\sigma}^{~~~\rho} \equiv \partial_\mu{\cal A}_{\nu~\sigma}^{~\,\rho}+{\cal A}_{\mu~\lambda}^{~\,\rho}{\cal A}_{\nu~\sigma}^{~\,\lambda}- (\mu\leftrightarrow \nu),$$
rather than just one.
These invariants are the following.
\begin{itemize}
\item The usual Ricci-like scalar
 ${\cal R}\equiv {\cal R}_{\mu\nu}^{~~~\mu\nu}$.
 \vspace{0.3cm}
 \item The parity-odd Holst invariant~\cite{Hojman:1980kv,Nelson:1980ph,Holst:1995pc} ${\cal R'}\equiv \epsilon^{\mu\nu\rho\sigma}{\cal R}_{\mu\nu\rho\sigma}/\sqrt{-g}$,
where $\epsilon^{\mu\nu\rho\sigma}$ is the Levi-Civita symbol ($\epsilon^{0123}=1$) and $g$ is the metric determinant. 
\end{itemize}

In GR, where ${\cal A}_{\mu~\sigma}^{~\,\rho}$ equals the Levi-Civita connection, ${\cal R}$ coincides with the Ricci scalar $R$, but ${\cal R'}=0$. Thus, in metric-affine gravity ${\cal R'}$ can be understood as a component of the connection.  
The key idea of Ref.~\citen{Salvio:2022suk} is to identify the inflaton with ${\cal R'}$.  To do so ${\cal R'}$ has to be dynamical and independent of the metric.

\section{The Minimal Model}

The simplest model that realizes the idea of the previous section features the following inflationary action
\be S_I= \int d^4x\sqrt{-g}\left( \alpha{\cal R}+\beta {\cal R'} + c {\cal R'}^2\right). \label{SI}\ee
Indeed, 
\begin{itemize}
\item for $c=0$ one can easily show, by solving the connection equations, that $S_I$ reduces to the pure Einstein-Hilbert action, having identified $\alpha=\bp^2/2$, with $\bp$ the reduced Planck mass; 
\item for $c\neq0$, standard auxiliary field methods show that an extra spin-0 parity odd dynamical field $\zeta'$ (the ``pseudoscalaron") is present and equals  ${\cal R'}$ on shell~\cite{Hecht:1996np,BeltranJimenez:2019hrm,Pradisi:2022nmh}; 
\item the $\beta {\cal R'}$ term, the ``Holst term", is necessary to obtain a suitable inflaton potential, as we will see; 
the quantity $\bp^2/(4\beta)$ is called the  Barbero-Immirzi parameter~\cite{Immirzi:1996di,Immirzi:1996dr}.
\end{itemize}

$S_I$ can be recast in the following metric form (where the connection equals the Levi-Civita one)
$$S_I = \int d^4x\sqrt{-g}\left[\frac{M_P^2}{2} R-\frac{(\partial \omega)^2}{2} -U(\zeta'(\omega)) \right],$$
with $U(\zeta')=c \zeta'^2$ ($c\geq 0$ for stability reasons) and 
\bea \zeta'(\omega) &=& \frac1{2c}\left(\frac{\bp^2 \tanh X(\omega)}{4\sqrt{1-\tanh^2X(\omega)}}-\beta\right), \nonumber \\ ~~X(\omega)&\equiv& \sqrt{\frac{2}{3}}\frac{\omega}{\bp}+\tanh ^{-1}\left(\frac{4 \beta }{\sqrt{16 \beta ^2+\bp^4}}\right).\nonumber
\eea
This model is thus a scalar-tensor theory without ghosts and tachyons.

\section{Inflation}	

What is remarkable about this construction is that the pseudoscalaron potential can drive inflation in agreement with data, as I illustrate in this section.

As we will see, this model admits a slow-roll description, {\it i.e.} 
\be \epsilon \equiv\frac{\bp^2}{2} \left(\frac{1}{U}\frac{dU}{d\omega}\right)^2\ll 1, \quad \eta \equiv \frac{\bp^2}{U} \frac{d^2U}{d\omega^2}\ll 1. \nonumber\ee
Thus, I use slow-roll formul\ae~for the number of e-folds $N_e$ as a function of  $\omega$  \be N_e(\omega) = N(\omega)  - N(\omega_{\rm end}),  \qquad N(\omega)  = \frac1{\bp^2}  \int^\omega d\omega'  \,  U\left(\frac{dU}{d\omega'}\right)^{-1} \nonumber  \ee
(the value of $\omega$ at the end of inflation, $\omega_{\rm end}$,
 satisfies $\epsilon(\omega_{\rm end}) = 1$) and for the scalar spectral index $n_s$, the tensor-to-scalar ratio  $r$ and the curvature power spectrum $P_R$
 $$ n_s = 1- 6\epsilon +2\eta, \quad r =16\epsilon, \quad P_R= \frac{U/ \epsilon}{24\pi^2 \bp^4}. $$
 
 There are analytic expressions not only for $\epsilon$, $\eta$, $n_s$, $r$ and $P_R$, but also for $N$ and $N_e$:
\bea \epsilon(\omega)&=& \frac{4 \bp^4 \cosh^2X(\omega)}{3 \left(\bp^2 \sinh X(\omega)-4 \beta \right)^2}, \nonumber\\ 
\eta(\omega) &=& \frac{4 \bp^2 \left(\bp^2 \cosh \left(2 X(\omega)\right)-4 \beta  \sinh X(\omega)\right)}{3 \left(\bp^2 \sinh X(\omega)-4 \beta \right)^2}, \nonumber
\\
N(\omega) &=&
 \frac{3}{4} \log \left(\cosh X(\omega)\right)-\frac{3 \beta  \arctan\left(\sinh X(\omega)\right)}{\bp^2},  \nonumber\\
n_s(\omega)&=& 1-\frac{8 \bp^4 \cosh ^2X(\omega)}{\left(\bp^2 \sinh X(\omega)-4 \beta \right)^2}+\frac{8 \bp^2 \left(\bp^2 \cosh \left(2X(\omega)\right)-4 \beta  \sinh X(\omega)\right)}{3 \left(\bp^2 \sinh X(\omega)-4 \beta \right)^2}, \nonumber \\
r(\omega)&=&\frac{64\bp^4\cosh^2 X(\omega)}{3 \left(\bp^2 \sinh X(\omega)-4 \beta \right)^2}, \nonumber
\\
P_R(\omega)&=&\frac{\left(\beta -\frac{1}{4} \bp^2 \sinh X(\omega)\right)^2 \left(\bp^2 \sinh X(\omega)-4 \beta \right)^2 \text{sech}^2X(\omega)}{128 \pi ^2 c \bp^8}. \nonumber
\eea
Moreover, the equation $\epsilon(\omega_{\rm end}) = 1$ can be solved for real $\omega_{\rm end}$ whenever $192 \beta^2\geq4\bp^2$ and one finds two solutions, which I call $\omega_\pm$:
\be \omega_\pm =\sqrt{\frac{3}{2}} \bp \left(\sinh^{-1}\left(\pm\sqrt{\frac{192 \beta ^2}{\bp^4}-4}-\frac{12 \beta }{\bp^2}\right)-\tanh ^{-1}\left(\frac{4 \beta }{\sqrt{16 \beta ^2+\bp^4}}\right)\right).\ee
Given the shape of $U(\zeta'(\omega))$ and the fact that $\omega_{\rm end}$ is the value of $\omega$ at the end of inflation, we take $|\omega_{\rm end}|=\min(|\omega_+|,|\omega_-|)$.

\begin{figure}[h!]
\begin{center} 
\hspace{-0.27cm} 
\vspace{0.4cm}
\includegraphics[scale=0.32]{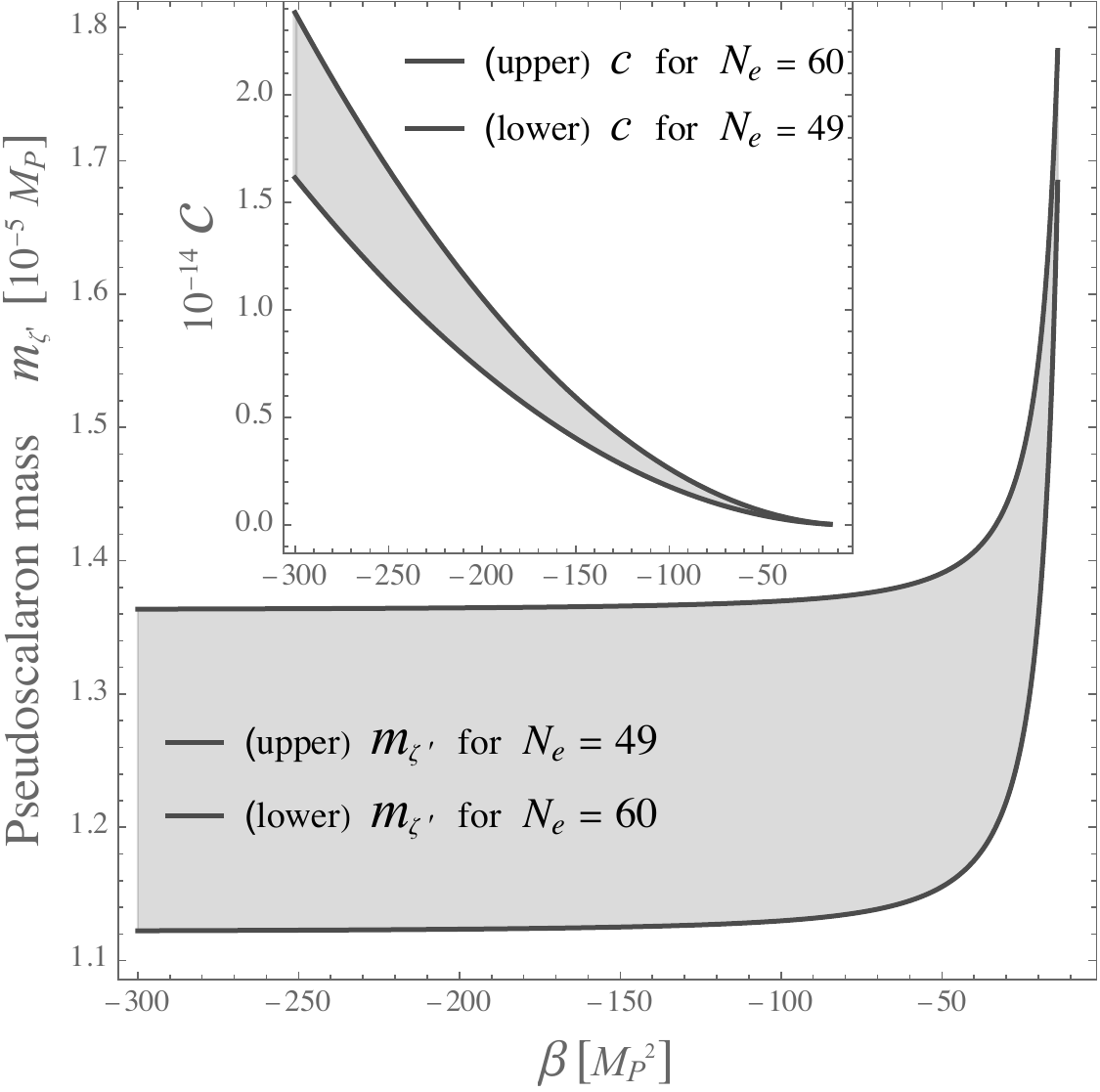}
\hspace{0.05cm}  \includegraphics[scale=0.32]{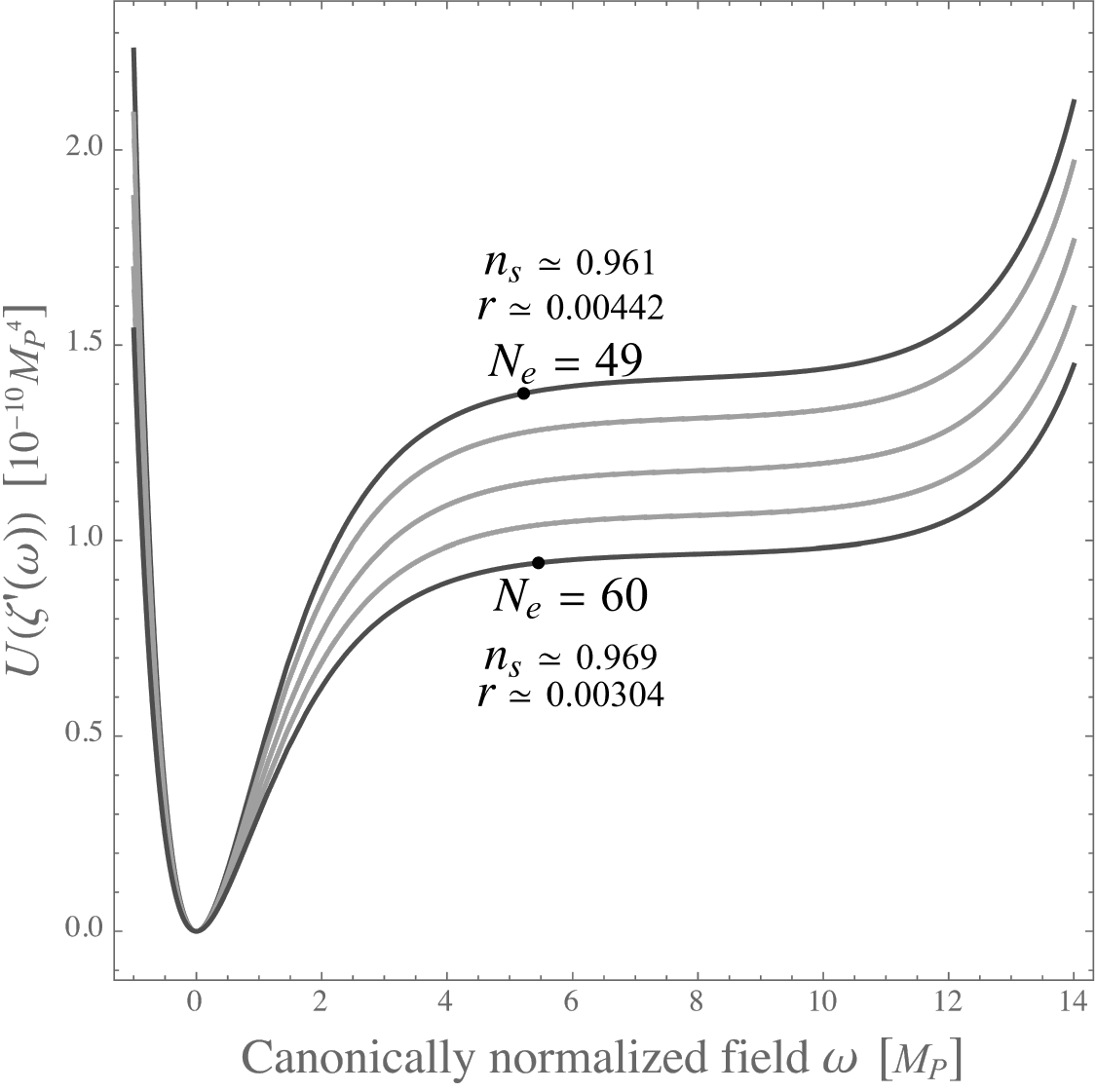}
\vspace{-0.6cm}
 \end{center} 
   \caption{ {\bf Left plot:} the pseudoscalaron mass and the corresponding value of $c$ (in the inset)  as a function of $\beta$. {\bf Right plot:} the corresponding pseudoscalaron potential for $\beta=-80\bp^2$. Also, the dots in the right plot correspond to the values of the inflaton for which $N_e=49$ (upper curve) and $N_e=60$ (lower curve) are realized; the corresponding predictions for $n_s$ and $r$ (in good agreement with the latest observational bounds) are displayed. }
\label{U-c}
\end{figure}

 In Fig.~\ref{U-c}  I show the potential of $\omega$  (right plot) and its  mass $m_\omega=m_{\zeta'}$ (left plot) by setting $c$ in a way that~\cite{Ade:2015lrj,Akrami:2018} $P_R=2.1 \times 10^{-9}$ at $N_e$ e-folds before the end of inflation. In the $\omega$ potential there is a plateau, which increases for larger $|\beta|$ and disappears as $\beta\to0$, which is why the $\beta {\cal R'}$ term in $S_I$ is necessary. In the right plot of Fig.~\ref{U-c} $\sqrt{|\beta|} = 4 \sqrt{5}\bp$ is chosen and is enough to even have     60 e-folds.

\begin{figure}[t]
\begin{center} 
\hspace{-0.37cm}
  \includegraphics[scale=0.31]{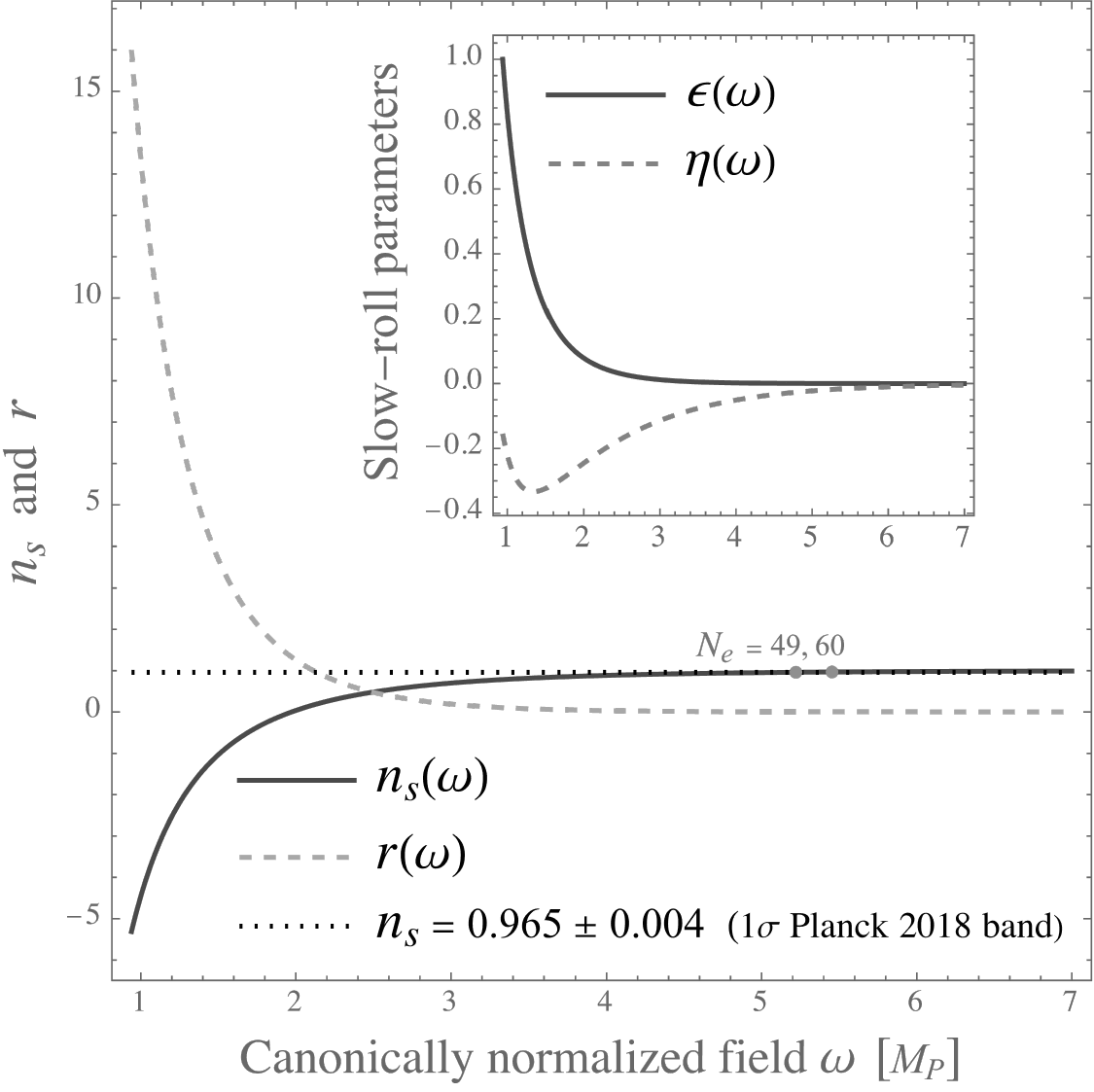} 
  \hspace{0.3cm} 
  \includegraphics[scale=0.32]{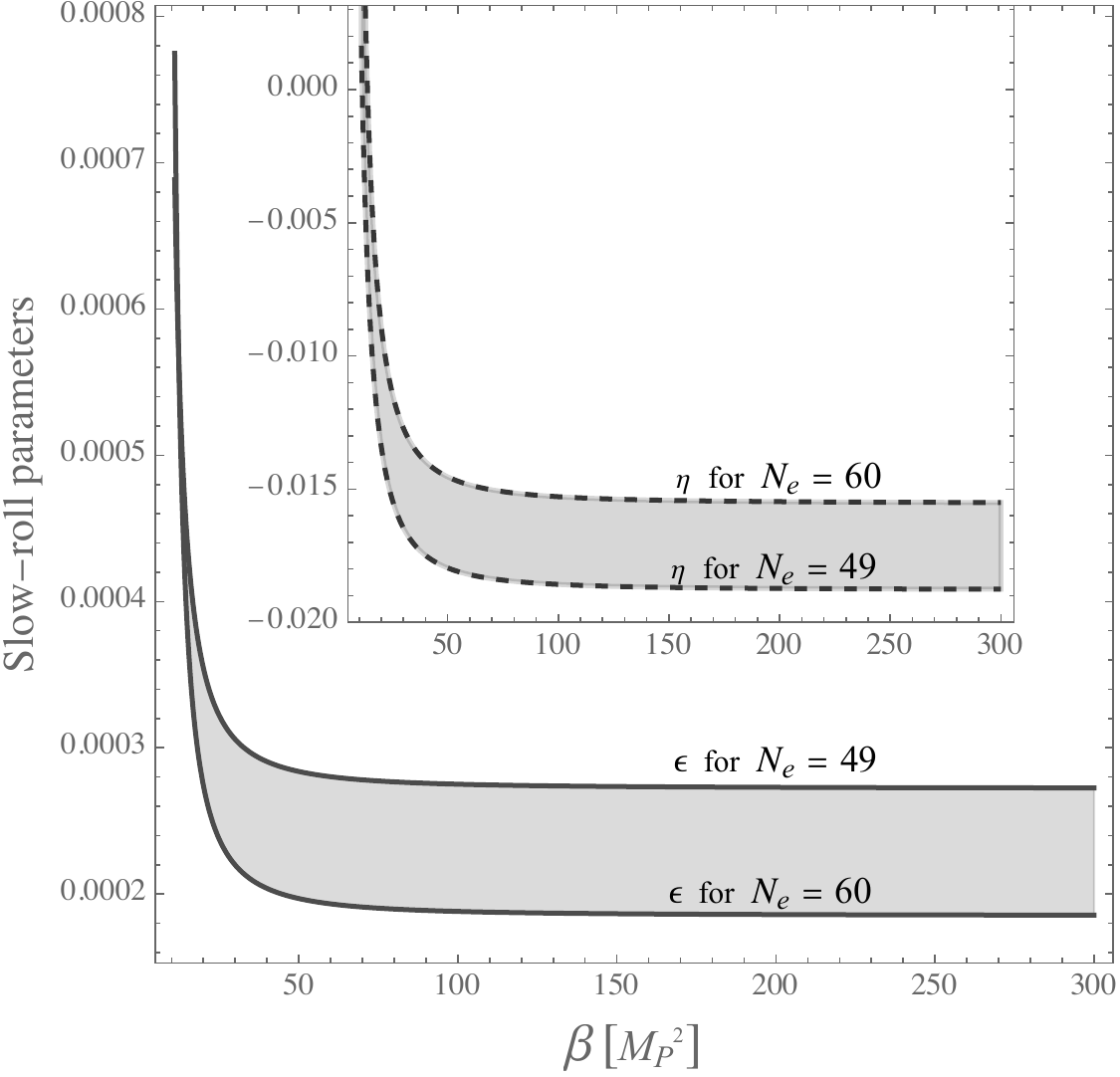} \\ 
 \vspace{0.016cm}
 \includegraphics[scale=0.31]{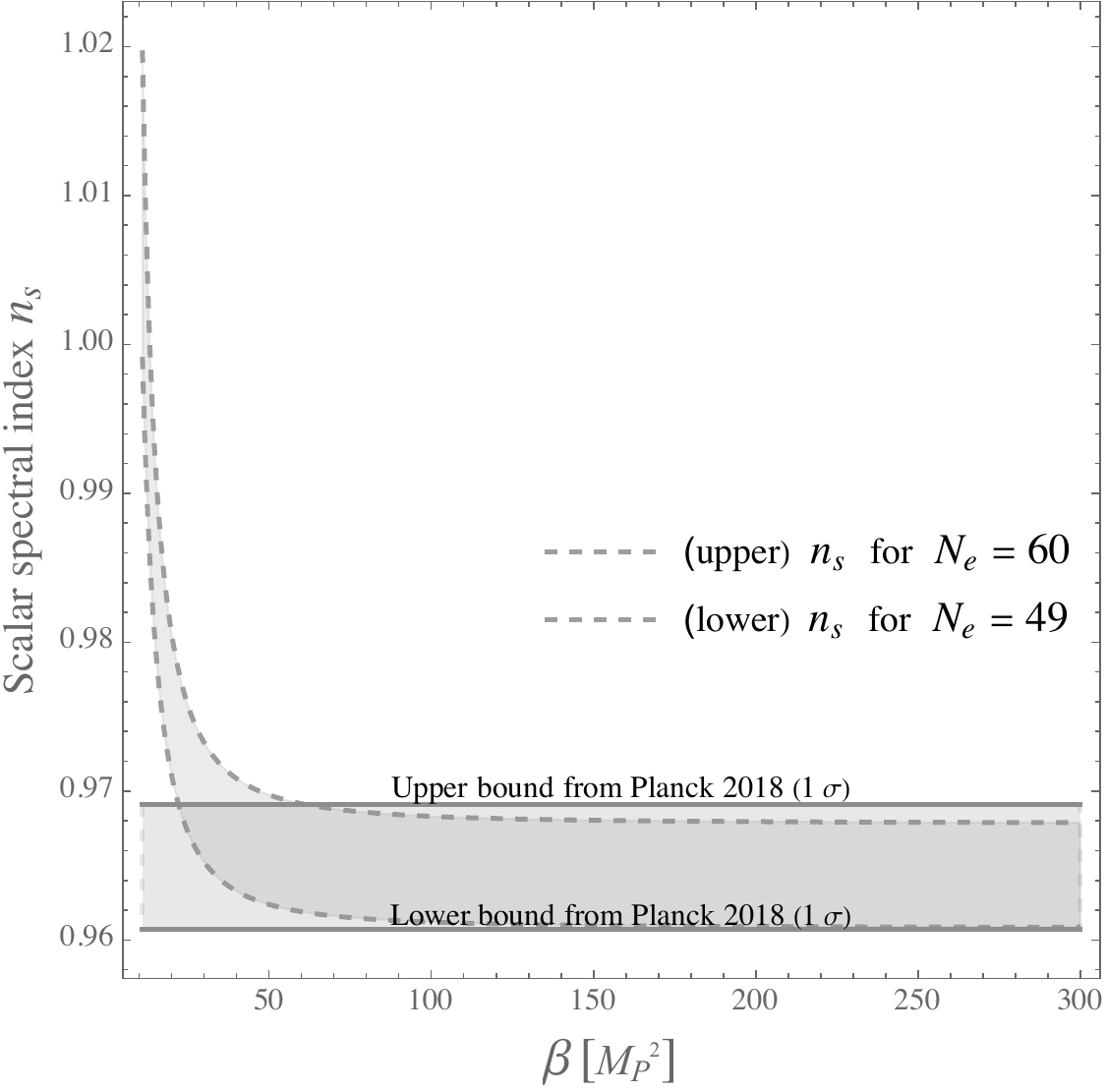}  
  \hspace{0.3cm} 
  \includegraphics[scale=0.31]{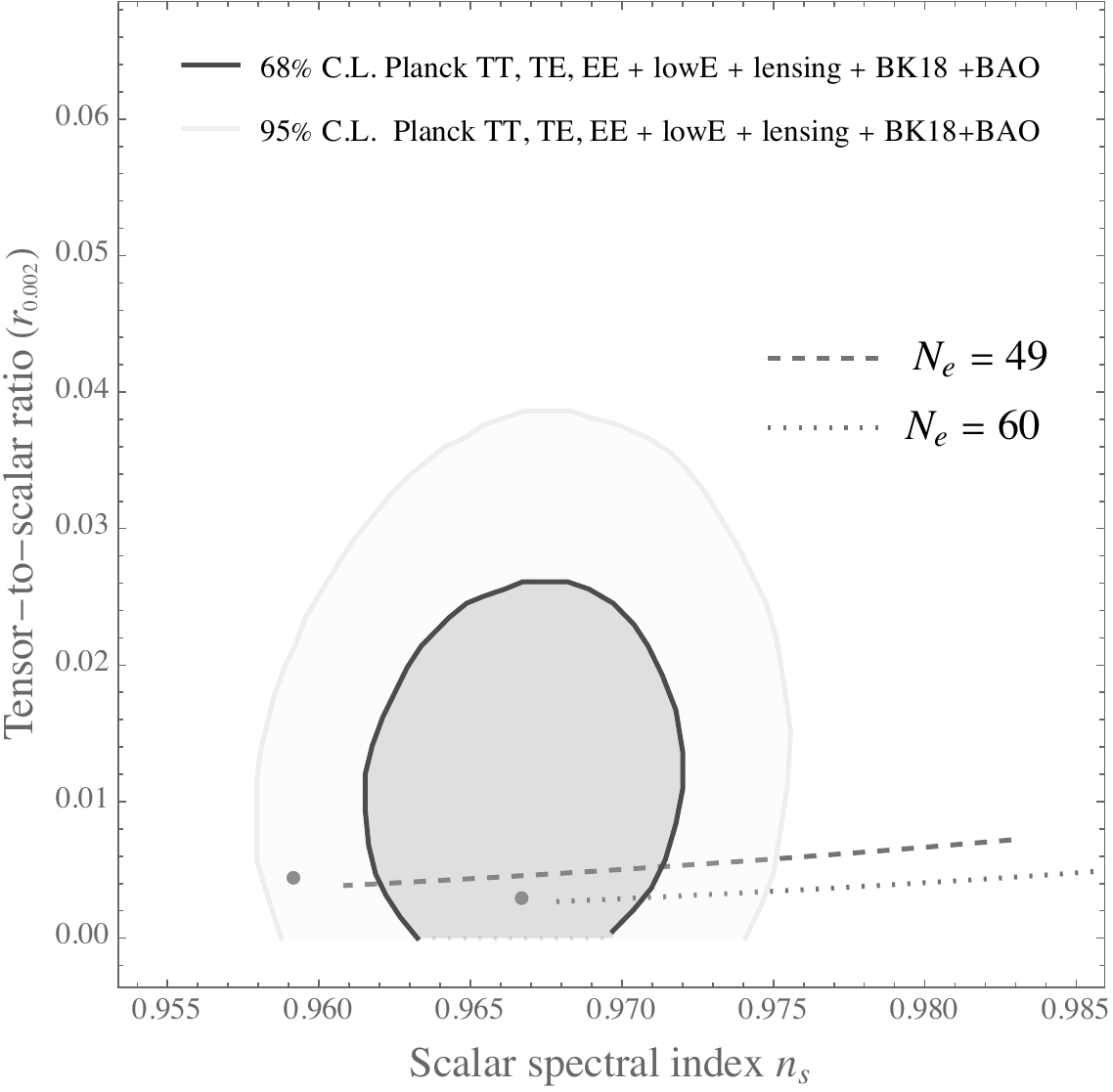}
 \end{center} 
   \caption{  In the top left plot the scalar spectral index   and the tensor-to-scalar ratio  as functions of the canonically normalized pseudoscalaron are displayed; in the inset the slow-roll parameters are given; there $\beta = -300 \bp^2$.
    In the other plots the slow-roll parameters,   the scalar spectral index   and the tensor-to-scalar ratio  as functions of $\beta$ are given. The two dots in the bottom right plot are the predictions of Starobinsky inflation.}
\label{inflationp}
\end{figure}

In Fig.~\ref{inflationp} it is shown  that slow-roll inflation not only occurs, but is also  remarkably compatible with the most recent CMB observations~\cite{Ade:2015lrj,Akrami:2018,BICEP:2021xfz} for large $|\beta|$ (i.e.~small values of the Barbero-Immirzi parameter) and for an appropriate value of $N_e$.  In Fig.~\ref{inflationp} I compare the observations and the theoretical predictions as functions of 
$\beta$ and show that viable slow-roll inflation with $N_e\simeq 49$ occurs already for $\sqrt{|\beta|} \gtrsim 4\bp$ and with $N_e\simeq 60$ for $\sqrt{|\beta|} \gtrsim 8\bp$.  In that figure $r_{0.002}$ is the value of $r$ at the reference momentum scale $0.002~{\rm Mpc}^{-1}$, used in the observations. In Fig.~\ref{inflationp} I also report the predictions of Starobinsky inflation~\cite{Starobinsky:1980te} for $n_s$ and $r$; the predictions of pseudoscalaron inflation approach (but do not  reach) those of Starobinsky inflation for $|\beta|\to \infty$, while for finite $\beta$ they differ significantly.

The future space mission LiteBIRD~\cite{LiteBIRD:2022cnt} will be able to test this scenario.

\section{Reheating}

 If $\omega$ decays into some Standard Model (SM) particles with width $\Gamma_\omega$ the reheating temperature $T_{\rm RH}$ is 
$$ T_{\rm RH}\gtrsim \min\left( \left(\frac{45 \Gamma^2_{\omega}\bp^2}{4\pi^3 g_*}\right)^{1/4},\left(\frac{30 \rho_{\rm vac}}{\pi^2 g_*}\right)^{1/4}\right), $$
where $g_*$ is the effective number of relativistic species at temperature $T_{\rm RH}$ and $\rho_{\rm vac}$ is the $\omega$ vacuum energy density (the full energy budget of the system).

\subsection{The pseudoscalaron decay into a fermion pair}
As a first example, consider a fermion $f$ represented by a Dirac spinor $\Psi$  with action ($e_a^\mu$ are the tetrads and ${\cal D}_\mu$ is the covariant derivative built with ${\cal A}_{\mu~\sigma}^{~\,\rho}$)
$$ S_f = \int \sqrt{-g} \, \frac12 \overline \Psi (i \gamma^a e_a^\mu {\cal D}_\mu-m_f)\Psi +{\rm h.c.}\, . $$
By using the connection equations with the formalism of Ref.~\citen{Pradisi:2022nmh}, one finds the following effective pseudoscalaron-fermion-fermion interaction 
 \be {\cal L}_{\omega ff} =   \frac{c_{\omega ff}}{\bp} \, \partial_\mu\omega\,\overline\Psi \gamma_5 \gamma^\mu\Psi, \qquad c_{\omega ff} \equiv \left[\frac{3\bp}{1+16 B^2}\frac{dB}{d\omega} \right]_{\omega=0} =\sqrt{\frac{3\bp^4}{8(\bp^4+16 \beta ^2)}},\nonumber \ee
 with $B =(\beta+2c\zeta'(\omega))/\bp^2$. This leads  to the decay $\omega\to ff$ with width
$$ \Gamma_{\omega\to ff} = |c_{\omega ff}|^2\frac{m_{\omega} m_f^2}{2\pi \bp^2} \sqrt{1-\frac{4m^2_f}{m_\omega^2}} $$
and can efficiently reheat the universe up to a temperature above the electroweak scale if the fermion mass $m_f$ is very large compared to that  scale.

It is possible to engineer a model where there is such a very heavy fermion with sizable couplings to SM particles.

\subsection{The pseudoscalaron decay into a scalar ({\it e.g.}~Higgs) pair}

In order to keep my analysis as model independent as possible, I consider here another channel: the decay of $\omega$ into two identical real scalar particles, {\it e.g.} two Higgs bosons (the Higgs is anyhow needed for any model to be viable).

This is possible when there is a non-minimal coupling between the real (canonically normalized) scalar field $\phi$ in question and ${\cal R}$  in the action: 
$$ S_{\rm nm}=\int \sqrt{-g}\frac{\xi \phi^2}{2} {\cal R}.$$
$S_{\rm nm}$ is known to be generated by quantum corrections and, so, it is more natural to include it. Solving the connection equations with the results of Ref.~\citen{Pradisi:2022nmh}, one finds
$$ {\cal L}_{\omega \phi\phi} =   \frac{c_{\omega \phi\phi}}{\bp} \, \partial_\mu\omega\,\phi \partial^\mu\phi, \qquad c_{\omega \phi\phi} = \left[\frac{48\xi \bp B}{1+16 B^2} \,\frac{dB}{d\omega}\right]_{\omega=0}=\frac{4 \sqrt{6} \beta  \xi }{\sqrt{\bp^4+16 \beta ^2}}. $$
${\cal L}_{\omega \phi\phi}$ only arises through the  Holst term  because $c_{\omega \phi\phi}\to 0$ as $\beta\to 0$ and gives  $$ \Gamma_{\omega\to \phi\phi} = |c_{\omega \phi\phi}|^2\frac{m_{\omega}^3}{16\pi \bp^2} \sqrt{1-\frac{4m^2_\phi}{m_\omega^2}}, $$
 where $\Gamma_{\omega\to \phi\phi}$ is the decay width of $\omega\to \phi\phi$ and  $m_\phi$ is the mass of $\phi$.
  The produced Higgs particles later decay into other SM particles, such as quarks and leptons. This channel can efficiently and naturally reheat the universe up to a temperature much above the electroweak scale, even when $\phi$ is the  Higgs, so  it does not  require any beyond-the-SM physics ({\it e.g.}~taking $m_\phi\ll m_\omega$, $g_*\sim 10^2$ and $\beta \gtrsim \bp^2$ one finds $T_{\rm RH}\gtrsim 10^9 |\xi|$~GeV).
  
\section{Conclusions}	

\begin{itemize} 
\item  It has been found that a pseudoscalar component of a dynamical connection, which is independent of the metric, can drive inflation in agreement with data. 

\item  This pseudoscalaron is  the parity odd Holst invariant and inflationary predictions in agreement with data have been found for small values of the Barbero-Immirzi parameter, where the inflaton potential forms a plateau.

\item The predictions approach, but do not reach, those of Starobinsky inflation as the Barbero-Immirzi parameter vanishes; instead, for finite values  the predictions differ significantly.

\item  Pseudoscalaron inflation can be tested by future CMB observations, {\it e.g.} those of LiteBIRD.

\item Lastly, the decays of the pseudoscalaron into Higgs particles can efficiently reheat the universe after inflation  to a high enough reheating temperature; the considered channel is made possible by the presence of an independent connection: the Holst term (which is needed for these decays to occur) would be absent if the full connection were exactly the Levi-Civita one. The reheating temperature could be further increased by other channels, such as decays into very massive fermions. 
\end{itemize}
As an outlook, it would be interesting to calculate other contributions to reheating. Moreover, it would also be interesting to engineer a fully scale invariant version of this model, perhaps along the lines of Refs.~\citen{Shaposhnikov:2008xi,Salvio:2014soa,DimTr}. Indeed, the  term, $c{\cal R'}^2$, is compatible with scale (and even Weyl) invariance, but the others in Eq.~(\ref{SI}) are not.

\section*{Acknowledgments}

I thank G.~Pradisi 
  for useful discussions. This work has been partially supported by the grant DyConn from the University of Rome Tor Vergata.
%
%
%


\begin{thebibliography}{0}    

\bibitem{Baldazzi:2021kaf}
A.~Baldazzi, O.~Melichev and R.~Percacci,
Annals Phys. \textbf{438} (2022), 168757.

\bibitem{Ade:2015lrj}
  P.~A.~R.~Ade {\it et al.} [Planck Collaboration],
  Astron.\ Astrophys.\  {\bf 594} (2016) A20.  

\bibitem{Akrami:2018}  Y.~Akrami {\it et al.} [Planck Collaboration],
  Astron. Astrophys. \textbf{641} (2020), A10. 
  
\bibitem{BICEP:2021xfz}
P.~A.~R.~Ade \textit{et al.} [BICEP and Keck],
Phys. Rev. Lett. \textbf{127} (2021) no.15, 151301. 

\bibitem{Salvio:2022suk}
A.~Salvio,
Phys. Rev. D \textbf{106} (2022) no.10, 103510.

\bibitem{DiMarco:2023ncs}
A.~Di Marco, E.~Orazi and G.~Pradisi,
[arXiv:2309.11345 [hep-th]].

\bibitem{Pradisi:2022nmh}
G.~Pradisi and A.~Salvio,
Eur. Phys. J. C \textbf{82} (2022) no.9, 840.

\bibitem{Hojman:1980kv}
R.~Hojman, C.~Mukku and W.~A.~Sayed,
Phys. Rev. D \textbf{22} (1980), 1915-1921.

\bibitem{Nelson:1980ph}
P.~C.~Nelson,
Phys. Lett. A \textbf{79} (1980), 285.

\bibitem{Holst:1995pc}
S.~Holst,
Phys. Rev. D \textbf{53} (1996), 5966-5969.


\bibitem{Hecht:1996np}
R.~D.~Hecht, J.~M.~Nester and V.~V.~Zhytnikov,
Phys. Lett. A \textbf{222} (1996), 37-42
 

\bibitem{BeltranJimenez:2019hrm}
J.~Beltr\'an Jim\'enez and F.~J.~Maldonado Torralba,
Eur. Phys. J. C \textbf{80} (2020) no.7, 611.

\bibitem{Immirzi:1996di}
G.~Immirzi,
Class. Quant. Grav. \textbf{14} (1997), L177-L181.

\bibitem{Immirzi:1996dr}
G.~Immirzi,
Nucl. Phys. B Proc. Suppl. \textbf{57} (1997), 65-72.




\bibitem{Starobinsky:1980te}
A.~A.~Starobinsky,
Phys. Lett. B \textbf{91} (1980), 99-102.

\bibitem{LiteBIRD:2022cnt}
E.~Allys \textit{et al.} [LiteBIRD],
[arXiv:2202.02773 [astro-ph.IM]].

\bibitem{Shaposhnikov:2008xi}
M.~Shaposhnikov and D.~Zenhausern,
Phys. Lett. B \textbf{671} (2009), 162-166.

\bibitem{Salvio:2014soa}
  A.~Salvio and A.~Strumia,
  JHEP {\bf 1406} (2014) 080. 

\bibitem{DimTr}A.~Salvio,
Int. J. Mod. Phys. A \textbf{36} (2021) no.08n09, 2130006.
  
%
%
%
%
%
%
%
%

\end{thebibliography}
\end{document}